\documentclass[twocolumn,showpacs,preprintnumbers,amsmath,amssymb]{revtex4}

\usepackage{graphicx}
\usepackage{dcolumn}
\usepackage{bm}

\def\bear{\begin{eqnarray}}
\def\ear{\end{eqnarray}}

\begin{document}

\title{Non-adiabatic Quantum Vlasov Equation for Schwinger Pair Production}

\author{Sang Pyo Kim}
\medskip
\affiliation{Department of Physics, Kunsan National University, Kunsan 573-701, Korea\\
Instituto de F\'{i}sica y Matem\'{a}ticas, Universidad Michoacana de San Nicol\'{a}s de Hidalgo, Apdo. Postal 2-82, C.P. 58040, Morelia, Michoac\'{a}n, Mexico}

\author{Christian Schubert}
\affiliation{Instituto de F\'{i}sica y Matem\'{a}ticas, Universidad Michoacana de San Nicol\'{a}s de Hidalgo, Apartado Postal 2-82, C.P. 58040, Morelia, Michoac\'{a}n, Mexico}

\medskip


\begin{abstract}
Using Lewis-Riesenfeld theory, we derive an exact non-adiabatic master equation describing the time evolution of the
QED Schwinger pair-production rate for a general time-varying electric field.
This equation can be written equivalently as a first-order matrix equation, as a
Vlasov type integral equation, or as a third-order differential equation. In the last version
it relates to the Korteweg-de Vries equation, which allows us to construct an exact solution
using the well-known one-soliton solution to that equation.
The case of time-like delta function pulse fields is also shortly considered.
\end{abstract}

\maketitle

\section{Introduction}

Vacuum pair production by a strong electric field, predicted by Schwinger in 1951 \cite{schwinger},
may now finally be seen due to the construction of ultra-strong laser systems \cite{QEDtest}.
However, the corresponding fields are very different from the few special configurations
for which an exact calculation of the pair creation rate is possible. Thus, recently there
has been increased interest in the development of approximation schemes, such as
semiclassical methods \cite{dunneschubert,kimpage,dumludunne} and Monte Carlo simulations \cite{montecarlo}.

A case that is relatively
amenable to an exact treatment is the one of a purely time-dependent electric field.
Here the spatial momentum is a good quantum number, which allows one to reduce the
time evolution of the system to a collection of mode equations labeled by the fixed
momentum $k$. The pair production calculation can then be further reduced to
a one-dimensional scattering problem, suitable for standard numerical or WKB methods \cite{breitz,popov,gitgav}.
Alternatively, the mode equation can be transformed to
the quantum Vlasov equation, an integral equation
for ${\cal N}_k(t)$, the total expected number of created pairs in the mode $k$
\cite{kescm,kme,rau,srsbtp,healgi} (see Ref. \cite{dumlu} for a comparison of the two approaches).

In this paper, we reconsider the time evolution of the QED Hamiltonian in
a time-varying field using Lewis-Riesenfeld invariant theory \cite{lewisriesenfeld} and a suitable operator
basis forming a spectrum generating algebra $SU(1,1)$. We derive an exact non-adiabatic master equation 
for the time evolution of the Schwinger pair-production rate.
This equation can be written equivalently as a first-order matrix equation, as a quantum
Vlasov equation, or as a third-order differential equation. For a specific solution
ansatz this third-order equation relates to the Korteweg-de Vries (KdV) equation, which allows us to construct an
exact solution using the well-known one-soliton solution to that equation.
We also consider the case of alternating time-like delta function pulse fields, a type of fields
which is of relevance for a recent proposal to apply Ramsey interferometry to the Schwinger effect
\cite{akkermansdunne}. 

\section{Derivation of the Master Equation}

We will give the derivation of the master equation for the scalar QED case; the derivation for
the spinor QED case is similar, and will be included in a forthcoming, more detailed publication
\cite{bigpaper}. A scalar particle with charge $q$ and mass $m$ in a
homogeneous time-dependent electric field with the gauge potential
$A_{\parallel} (t)$ has the Fourier decomposed Hamiltonian of time-dependent oscillators
[in units of $\hbar = c = 1$]
\begin{eqnarray}
\hat H (t) = \int \frac{d^3 k}{(2 \pi)^3} \Bigl[\pi_k^{\dagger} \pi_k + \omega^2_k (t) \phi_k^{\dagger} \phi_k \Bigr],
\label{ham}
\end{eqnarray}
where
\begin{eqnarray}
\omega_k^2 (t) &=& (k_{\parallel} - qA_{\parallel} (t))^2 + {\bf k}_{\perp}^2 + m^2.
\end{eqnarray}
We will quantize the theory in the Schr\"{o}dinger picture, where
the time-dependent quantum state obeys the functional Schr\"{o}dinger equation
\begin{eqnarray}
i\frac{\partial \Psi (t)}{\partial t} = \hat{H} (t) \Psi (t).
\label{schroed}
\end{eqnarray}
In this picture the field operators $\hat{\phi} (x)$ and $\hat{\pi}(x)=\hat{\dot{\phi}}^{\dagger}$ are time-independent with
the momentum space commutation relations
\begin{eqnarray}
[\hat\phi_k,\hat\pi_{k'}] = [\hat\phi_k^{\dagger},\hat\pi_{k'}^{\dagger}] = i(2 \pi)^3 \delta_{kk'},
\end{eqnarray}
but the corresponding creation and annihilation operators with the equal-time commutators
\begin{eqnarray}
\lbrack\hat{a}_{k} (t), \hat{a}^{\dagger}_{k'} (t)] = [\hat{b}_{-k} (t), \hat{b}^{\dagger}_{-k'} (t)] = (2 \pi)^3 \delta_{kk'}
 \label{etcomm}
 \end{eqnarray}
are generally time-dependent \cite{mmt,cdms,ffvv,kim-lee},
\begin{eqnarray}
\hat{\phi}_k &=&  \hat{a}_k (t) \varphi_k (t) + \hat{b}^{\dagger}_{-k} (t) \varphi^*_k (t), \nonumber\\
\hat{\pi}_k &=&  \hat{a}^{\dagger}_k (t) \dot{\varphi}_k^* (t)
 + \hat{b}_{-k} (t) \dot{\varphi}_k (t).
 \label{defhatphipi}
\end{eqnarray}
Here $\varphi_k$ is an auxiliary field satisfying the classical mode equation
\begin{eqnarray}
\ddot{\varphi}_k (t) + \omega^2_k (t) \varphi_k (t) = 0, \label{mod eq}
\end{eqnarray}
as well as the Wronskian constraint
\begin{eqnarray}
{\rm Wr} [\varphi_k, \varphi^*_k] \equiv \varphi_k (t) \dot{\varphi}^*_k (t) - \varphi^*_k (t) \dot{\varphi}_k (t) = i.
\label{wrosnk}
\end{eqnarray}
The Eq. (\ref{mod eq}) and the Wronskian determine $\varphi_k (t)$ up to a phase factor,
which we fix by requiring that $\varphi_k (t) $ be real at the initial time $t_0$. Thus if
$t_0$ is finite, then for $t\leq t_0$ one has
\begin{eqnarray}
\varphi_{{\rm}k} (t) = \frac{e^{- i \omega_k (0)  (t-t_0)}}{\sqrt{2 \omega_k (0) }}
\label{phiasymp}
\end{eqnarray}
(for $t_0 = -\infty$ this should hold in the asymptotic sense).
We note that the operators $\hat{a}_{k} (t),\hat{b}_{-k} (t)$ and their hermitian
conjugates are Lewis-Riesenfeld invariants, that is, they fulfill the Liouville-von Neumann equation
\begin{eqnarray}
i \frac{\partial \hat{I}_k (t)}{\partial t}   + [\hat{I}_k (t), \hat{H}_k (t)] = 0,
\end{eqnarray}
as can be easily checked.

The ground state $\vert 0_k, t \rangle$ for the $k$-th mode is annihilated
by both $\hat{a}_{k} (t)$ and $\hat{b}_{-k} (t)$ and
the $n$-th excited state is
\begin{eqnarray}
\vert n_k, t \rangle := \frac{[\hat{a}^{\dagger}_k (t) \hat{b}^{\dagger}_{-k} (t)]^{n_k}}{n_k!}\vert 0_k, t \rangle\, .
\end{eqnarray}
Thus the total time-dependent vacuum state is given by
\begin{eqnarray}
\vert 0, t \rangle = \prod_k \vert 0_k, t \rangle.
\end{eqnarray}
In the free theory, the time-dependent vacuum state reduces to the Minkowski vacuum, as expected.
The scalar product for the quantized fields and their hermitian conjugates allows us to find the
Bogoliubov transformation between the past time $t_0$ and the present time $t$, which is given by
\begin{eqnarray}
\hat{a}_k (t_0) &=& \mu_k (t_0, t) \hat{a}_k (t) + \nu_k (t_0, t) \hat{b}_{- k}^{\dagger} (t), \nonumber\\
\hat{b}^{\dagger}_{-k} (t_0) &=& \mu^*_k (t_0, t) \hat{b}^{\dagger}_{-k} (t) + \nu^*_k (t_0, t) \hat{a}_{k} (t),\label{bog tr}
\end{eqnarray}
where
\begin{eqnarray}
\mu_k (t_0, t) &=& i {\rm Wr} [\varphi^*_k (t_0), \varphi_k (t)], \quad \nonumber\\
\nu_k (t_0, t) &=& i {\rm Wr} [\varphi^*_k (t_0), \varphi^*_k (t)]. \label{bog co}
\end{eqnarray}
The Bogoliubov coefficients satisfy the relation for bosons $|\mu_k (t_0, t)|^2 - |\nu_k (t_0, t)|^2 = 1$.
Our main object of interest, the
mean number of pairs present at time $t$ assuming that this number was $n_k$ at the initial time $t_0$, can now be read off from
\begin{eqnarray}
\langle n_k, t\vert \hat{a}^{\dagger}_k (t_0) \hat{a}_k (t_0) \vert n_k, t \rangle
 = |\nu_k (t_0, t)|^2 (2n_k +1) + n_k. \nonumber\\ \label{mean num}
\end{eqnarray}
Thus 

\bear
{\cal N}_k (t) := |\nu_k (t_0, t)|^2(2n_k +1)
\label{defN}
\ear
is the number of pairs
spontaneously produced from the initial vacuum by the electric field.

To obtain a time evolution equation for this quantity, we observe that
the time-dependent Hamiltonian (\ref{ham}) has the spectrum generating algebra $SU(1,1)$.
Choosing the Hermitian basis
\begin{eqnarray}
\hat{\cal M}_k^{(0)} (t_0) &=& \frac{1}{(2\pi)^3}\bigl\lbrack
 \hat{a}^{\dagger}_k (t_0) \hat{a}_k (t_0) + \hat{b}_{-k} (t_0)
\hat{b}^{\dagger}_{-k} (t_0)\bigr\rbrack, \nonumber\\
\hat{\cal M}_k^{(+)} (t_0) &=&
 \frac{1}{(2\pi)^3}\bigl\lbrack\hat{a}_k (t_0) \hat{b}_{-k} (t_0) + \hat{a}^{\dagger}_{k} (t_0)
\hat{b}^{\dagger}_{-k} (t_0)\bigr\rbrack, \nonumber\\
\hat{\cal M}_k^{(-)} (t_0) &=&   \frac{i}{(2\pi)^3}\bigl\lbrack \hat{a}_k (t_0) \hat{b}_{-k} (t_0) - \hat{a}^{\dagger}_{k} (t_0)
\hat{b}^{\dagger}_{-k} (t_0) \bigr], \nonumber\\
\label{her ba}
\end{eqnarray}
this algebra becomes
\begin{eqnarray}
\bigl[\hat{\cal M}_k^{(0)} (t_0), \hat{\cal M}_k^{(\pm)} (t_0) \bigr] &=& \pm 2i \hat{\cal M}_k^{(\mp)} (t_0), \nonumber\\
\quad \bigl[\hat{\cal M}_k^{(+)} (t_0), \hat{\cal M}_k^{(-)} (t_0) \bigr] &=& - 2i \hat{\cal M}_k^{(0)} (t_0).
 \label{spec al}
\end{eqnarray}
The correlators are the expectation values of Eq. (\ref{her ba}) with respect to $\vert n_k, t \rangle$, that is, of the number of produced pairs
and of pair creation and annihilation:
\begin{eqnarray}
1+2{\cal N}_k (t) &=& (2 |\nu_k (t_0, t)|^2 +1)(2n_k+1), \nonumber\\
{\cal M}^{(+)}_k (t) &=& (\mu_k (t_0, t) \nu_k (t_0, t) + \mu^*_k (t_0, t) \nu^*_k (t_0, t))\nonumber\\
&& \times (2n_k+1), \nonumber\\
{\cal M}^{(-)}_k (t) &=& i(\mu_k (t_0, t) \nu_k (t_0, t) - \mu^*_k (t_0, t) \nu^*_k (t_0, t))\nonumber\\
&& \times (2n_k +1). \label{cor}
\end{eqnarray}
Note that all three correlators are real and proportional to the quantum number $2n_k +1$, and thus
proportional to the ones defined by the vacuum state.

Using Eq. (\ref{bog co}) and the mode equation (\ref{mod eq}), we find the first order master equation
\begin{eqnarray}
\frac{d}{dt} \begin{pmatrix}
1+2{\cal N}_k \\
{\cal M}^{(-)}_k\\
{\cal M}^{(+)}_k
\end{pmatrix} = \begin{pmatrix}
  0 &  \Omega^{(-)}_k & 0 \\
     \Omega^{(-)}_k & 0 &  \Omega^{(+)}_k\\
     0& - \Omega^{(+)}_k &0 \end{pmatrix} \begin{pmatrix}
1+2{\cal N}_k \\
{\cal M}^{(-)}_k\\
{\cal M}^{(+)}_k
\end{pmatrix},\nonumber\\ \label{master eq}
\end{eqnarray}
where
\begin{eqnarray}
\Omega^{(\pm)}_k (t) := \frac{\omega_k^2 (t) \pm \omega_k^2 (t_0)}{\omega_k (t_0)}\, ,
\end{eqnarray}
with the initial conditions ${\cal N}_k = n_k$, ${\cal M}^{(\pm)}_k = 0$ at $t=t_0$ (where
$t_0$ may be $-\infty$). An immediate consequence of the master equation (\ref{master eq}) is the conservation of the quantity
\begin{eqnarray}
(1+2 {\cal N}_k )^2 - ({\cal M}^{(+)}_k)^2 - ({\cal M}^{(-)}_k)^2 = (1+2n_k )^2. \label{con law}
\end{eqnarray}
This relates to the conservation of charge, as well as to the invariance of the
Casimir operator for the $SU(1,1)$ algebra \cite{bigpaper}. The spinor QED case can be treated analogously \cite{bigpaper}.
As far as the master formula (\ref{master eq}) is concerned, the generalization to the
fermionic case requires only changing $1+2 {\cal N}_k (t)$ to
$1-2 {\cal N}_k (t)$, and replacing $\omega_k^2(t)$ by
\begin{eqnarray}
\omega_k^2 (t) = (k_{\parallel} -q A_{\parallel} (t))^2 + i qE(t)  + {\bf k}_{\perp}^2 + m^2.
\label{omegaspin}
\end{eqnarray}

\section{Alternative Formulations of the Master Equation}

The first order matrix equation (\ref{master eq}) can be equivalently rewritten both as a
single integral equation and as a third order linear differential equation.
Since we work with a fixed mode $k$, in this section we will generally
suppress the index $k$ and abbreviate $\omega_0  := \omega(t_0)$.
We will now also set $n_k=0$.

First, we combine the equations for ${\cal M}^{(\pm)}$ to a second order inhomogeneous equation for ${\cal M}^{(-)}$,
\begin{eqnarray}
\frac{d^2 {\cal M}^{(-)}}{dt^2} - \frac{\dot{\Omega}^{(+)}}{\Omega^{(+)}}
\frac{d {\cal M}^{(-)}}{dt} + (\Omega^{(+)})^2 {\cal M}^{(-)}
 \nonumber\\ =  \Omega^{(+)} \frac{d}{dt}
\Bigl[\frac{\Omega^{(-)}}{\Omega^{(+)}} (1+2 {\cal N}) \Bigr].
\label{m eq}
\end{eqnarray}
The homogeneous part of Eq. (\ref{m eq}) has the exact solutions
\begin{eqnarray}
{\cal M}^{(-)} (t) = C^{\pm} e^{\pm i \int_{t_0}^{t} dt' \Omega^{(+)} (t')}
\end{eqnarray}
with integration constants $C^{\pm}$. Using those in the usual way
to construct the solution of the inhomogeneous equation with the appropriate initial conditions, we
obtain the quantum Vlasov equation as the integral equation
\begin{eqnarray}
\frac{d}{dt} (1+ 2 {\cal N} (t))
&=& \Omega^{(-)} (t) \int_{t_0}^{t} dt' \Bigl[ \Omega^{(-)}(t')
(1+2 {\cal N} (t') ) \nonumber\\
&&\times  \cos( \int_{t'}^{t} dt'' \Omega^{(+)} (t'') ) \Bigr]. \label{new qve}
\end{eqnarray}

Second, inspection of the master equation (\ref{master eq}) shows, that its general solution can
be parameterized by a function $f(t)$ fulfilling the integral equation
\begin{eqnarray}
\dot f(t) = \frac{\Omega^{(-)}(t)}{\omega_0}
-2 \int_{t_0}^t dt' f(t') \bigl( \omega^2(t)+\omega^2(t')\bigr)
\label{f}
\end{eqnarray}
with the initial condition $f(t_0)=\dot f(t_0)=0$. Then, the correlators are given by
\begin{eqnarray}
1+2{\cal N} &=& 1 + \omega_0 \int_{t_0}^t dt' f(t') \Omega^{(-)}(t'), \nonumber\\
{\cal M}^{(-)} &=& \omega_0 f(t), \nonumber\\
{\cal M}^{(+)} &=& - \omega_0 \int_{t_0}^t dt' f(t') \Omega^{(+)}(t'). \label{solv}
\end{eqnarray}
Alternatively the integral equation (\ref{f}) can, taking one derivative, be converted into a third order linear differential equation,
\begin{eqnarray}
\dddot F + 4\omega^2\dot F+2{(\omega^2)}\dot{\phantom{.}} F
 =\frac{(\omega^2)\dot{\phantom{.}}}{\omega_0^2}
\label{DGL}
\end{eqnarray}
where
\begin{eqnarray}
F(t) := \int_{t_0}^t dt' f(t')
\end{eqnarray}
and the initial conditions are $F(t_0)=\dot F(t_0) = \ddot F(t_0)=0$.
Observe that $\ddot F$ is absent in Eq. (\ref{DGL}), which by Abel's theorem implies
that the Wronskian of the solutions of the corresponding homogeneous equation is constant.

The differential equation (\ref{DGL}) bears an interesting relationship to the
KdV equation. The form of the integral equation (\ref{f}) suggests the ansatz
\begin{eqnarray}
f(t)  =  \frac{(\omega^2)\dot{\phantom{.}}(t)}{8\omega_0^4}, \quad F(t)  =  \frac{\omega^2(t)-\omega_0^2}{8\omega_0^4}.
\label{ansatzf}
\end{eqnarray}
Defining $r(t) := \omega^2(t)/\omega_0^2$ and then $u(x,t) := -r(x-10t)$, one can show that $u$ satisfies the KdV equation,
\begin{eqnarray}
u_{xxx} - 6 uu_x + u_t = 0.
\label{kdv}
\end{eqnarray}
Thus we can use certain solutions of the KdV equation to calculate pair creation rates
for the corresponding electric fields.

\section{Exactly Solvable Cases}

We will now study two exactly solvable cases.
First, we consider the following soliton-type solution of the KdV equation
(see, e.g., Refs. \cite{dasbook,drazin-johnson,cksbook})
\begin{eqnarray}
u(x,t) = -1-\frac{2}{{\rm cosh}^2(x-10t)},
\label{usoliton}
\end{eqnarray}
which corresponds to
\begin{eqnarray}
r(t) = 1 + \frac{2}{\cosh^2(\omega_0 t)}, \quad F(t) =  \frac{1}{4\omega_0^2\cosh^2(\omega_0 t)}.
\label{rsoliton}
\end{eqnarray}
This is a solution to Eq. (\ref{DGL}) with the appropriate boundary conditions at $t_0 = -\infty$.
The gauge potential is
\begin{eqnarray}
qA(t) = k_{\parallel} - \sqrt{k_{\parallel}^2+ \frac{2\omega_0^2}{\cosh^2(\omega_0 t)}}.
\label{Asol}
\end{eqnarray}
From Eq. (\ref{solv}) we get the exact pair creation rate,
\begin{eqnarray}
{\cal N}(t) = \frac{1}{8\cosh^4(\omega_0 t)}.
\label{2Nsoliton}
\end{eqnarray}

Note that ${\cal N}(t)$ returns to zero for $t\to\infty$, which is due to the
solitonic character that makes the scattering reflectionless. In fact,
the mode solution to Eq. (\ref{mod eq}) is given by
\begin{eqnarray}
\varphi (t) = \frac{e^{- i \omega_0 t}}{\sqrt{2 \omega_0}} A(t),
\label{phisol}
\end{eqnarray}
where the amplitude is
\begin{eqnarray}
A(t) = (e^{2 \omega_0 t} +1)^{2} {}_2F_1(2, 2-i; 1-i; - e^{2 \omega_0 t})
\label{ampsol}
\end{eqnarray}
 with ${}_2F_1$ the hypergeometric function, and it does not have a negative frequency part in the future.
The Bogoliubov coefficient (\ref{bog co}) is
\begin{eqnarray}
\nu (t) = \frac{e^{2i \omega_0 t}}{2 \omega_0} \dot{A}^* (t),
\label{bogsol}
\end{eqnarray}
which approximately leads to ${\cal N} (t) = 2 e^{ 4 \omega_0 t}$ for $\omega_0 t \ll -1$
and ${\cal N} (t) = 2 e^{ - 4 \omega_0 t}$ for $\omega_0 t \gg 1$, the leading approximation to the exact formula (\ref{2Nsoliton}).
Thus there is no pair creation in this case, contrary to the somewhat similarly-looking
Sauter field case \cite{sauter}. This example also shows clearly that, as emphasized in Ref. \cite{healgi}, no direct physical
meaning should be ascribed to ${\cal N}_k(t)$ at intermediate times.

Second, we consider an electric field consisting of two opposite delta function pulses,
\begin{eqnarray}
E(t) = E_0 \delta (t) - E_0 \delta (t - t_1),
\end{eqnarray}
which has the gauge potential of a potential well \cite{akkermansdunne}: $A_{\parallel} = 0$ for $t < 0$ and $t> t_1$,
corresponding to $\omega_k(0)$,
and $A_{\parallel} = - E_0$ for $0 < t < t_1$, corresponding to $\omega_k$.
The master equation (\ref{master eq}) together with continuity at $t =0$ leads to
the pair production for the period $0 < t <t_1$
\begin{eqnarray}
1+ 2{\cal N}_k (t) &=& (2n_k +1) \Bigl(\frac{\Omega_k^{(+)}}{2 \omega_k} \Bigr)^2 \nonumber\\
&& \times \Bigl[1 - \Bigl(\frac{\Omega_k^{(-)}}{\Omega^{(+)}_k} \Bigr)^2  \cos(2 \omega_k t) \Bigr],
\label{resdelta}
\end{eqnarray}
and for the period $t > t_1$ it now remains constant, retaining its value for $t_1$.
Note that for a single delta function pulse ${\cal N}_k(t)$ keeps oscillating, so that
the limit $t\to\infty$ cannot be defined. This is presumably due
a combination of the unphysical character of such a field and the non-Markovian nature of the time evolution.

\section{Discussion and Conclusions}

The central results of this paper are the master
equation (\ref{master eq}) and associated quantum Vlasov equation (\ref{new qve}),
each describing the exact time evolution of the cumulative pair creation variable
${\cal N}_k(t)$ for an electric field that depends only on time, but is arbitrary otherwise. To the best of our
knowledge, these equations are new. 
We have concentrated here on scalar QED, leaving the details of the spinor QED case
to a more extensive publication \cite{bigpaper}.

In future work, we also plan to study the precise conditions under which a non-adiabatic
treatment is really necessary. To define the adiabatic approximation, we
write the mode solution in terms of the adiabatic basis \cite{kme}

\begin{eqnarray}
\varphi_k (t) = \alpha_k (t) \frac{e^{-i \theta (t)}}{\sqrt{2 \omega_k (t)}} + \beta_k (t)
\frac{e^{i \theta (t)}}{\sqrt{2 \omega_k (t)}},
\end{eqnarray}
where $\theta (t) = \int_{t_0}^t dt' \omega_k (t') $ and the Bogoliubov relation 
$|\alpha_k|^2 - |\beta_k|^2 = 1$ holds. We then replace $|{\nu_k(t_0,t)}|^2$ 
by $|{\beta_k}(t)|^2$ in the definition (\ref{defN}) of ${\cal N}(t)$.

From Eq. (\ref{bog co}) one can easily show that for this approximation to hold it is
sufficient to assume that

\begin{eqnarray}
\Bigl| \sqrt{\frac{\omega_k (t)}{\omega_k (0)}} - \sqrt{\frac{\omega_k (0)}{\omega_k (t)}} \Bigr|, \quad \Bigl|
\frac{\dot{\omega}_k (t)}{\omega_k^2 (t)} \Bigr| \ll |\beta_k (t)|
\label{ad con}
\end{eqnarray}
throughout the time evolution.
This criterium is similar, although not strictly equivalent,
to the one given in \cite{kme},

\bear 
\frac{\dot\omega}{\omega^2} \ll1, \quad \frac{\ddot \omega}{\omega^3} \ll 1.
\label{smallder}
\ear
In any case, all the inequalities in (\ref{ad con}),(\ref{smallder}) are certainly 
fulfilled for even the strongest laser sources which are presently existing or in
development. Those have a field strength still
much lower than the critical strength $E_c = m^2/e$ and the characteristic time scale much longer than the Compton time \cite{QEDtest}.

Concerning the relation of the master equation to the KdV equation, although there is a well-known
connection between the latter equation and one-dimensional quantum mechanical scattering
(see, e.g., Refs. \cite{dasbook,drazin-johnson,cksbook,feinberg}), it appears not to have been previously applied to the
Schwinger pair creation problem. It will be interesting to see whether also the
multi-soliton solutions of the KdV equation may be used in this context.

Finally, let us mention that it is straightforward to extend our master equation to the case of an initial state
which is a thermal state at temperature $T$.
As will be shown in Ref. \cite{bigpaper}, such a change leads
again only to an overall factor $\bigl(\coth (\beta \omega_k (0)/2) +1\bigr)$ multiplying
all three correlators $1+2{\cal N}_k,{\cal M}^{(\pm)}_k$, so that the master equation itself remains unaffected.

\acknowledgments
C.~S. thanks G.~V.~Dunne, J.~Feinberg and S.~P.~Gavrilov for helpful discussions.
The work of S.~P.~K. was supported in part by Basic Science Research Program through
the National Research Foundation of Korea (NRF) funded by the Ministry of Education, Science and Technology (2011-0002-520)
and the work of C.~S. was supported by CONACYT through Grant Conacyt Ciencias Basicas 2008 101353.

\end{document}